\documentclass[twocolumn,showpacs,preprintnumbers,amsmath,amssymb,prl,superscriptaddress]{revtex4}
\usepackage{graphicx}
\usepackage{mciteplus}

\begin{document}

\title{Bistability signatures in nonequilibrium charge transport through molecular quantum dots}

\author{Klaus Ferdinand Albrecht}
\affiliation{Physikalisches Institut, Albert--Ludwigs--Universit\"at Freiburg,
Hermann-Herder-Str. 3, D--79104 Freiburg, Germany}
\author{Haobin Wang}
\affiliation{Department of Chemistry and Biochemistry, MSC 3C,
New Mexico State University, Las Cruces, New Mexico 88003, USA}
\author{Lothar M\"uhlbacher}
\affiliation{Physikalisches Institut, Albert--Ludwigs--Universit\"at Freiburg,
Hermann-Herder-Str. 3, D--79104 Freiburg, Germany}
\author{Michael Thoss}
\affiliation{Institut f\"ur Theoretische Physik und
Interdisziplin\"ares Zentrum f\"ur Molekulare Materialien,
Friedrich-Alexander-Universit\"at Erlangen-N\"urnberg,
Staudtstr.\ 7/B2, D--91058 Erlangen, Germany}
\author{Andreas Komnik}
\affiliation{Institut f\"ur Theoretische Physik, Universit\"at Heidelberg,
D--69120 Heidelberg, Germany}

\date{\today}

\begin{abstract}

  We investigate the transient nonequilibrium dynamics of a molecular junction
  biased by a finite voltage and strongly coupled to internal vibrational
  degrees of freedom. Using two different, numerical exact techniques,
  diagrammatic Monte Carlo and the multilayer multiconfiguration time-dependent
  Hartree method, we show that the steady state current through the junction
  may depend sensitively on the initial preparation of the system, thus
  revealing signatures of bistability. The influence of the bias voltage and
  the transient dynamics on the phenomenon of bistability is
  analyzed. Furthermore, a possible relation to the phenomenon of stochastic
  switching in nanoelectromecanical devices is discussed.

\end{abstract}


\maketitle

Since the first realization of a single-molecule junction
the field of molecular electronics has seen
a rapid development \cite{cuniberti,Cuevas10}. Experimental investigations of
the conductance properties of single-molecule junctions have revealed a wealth
of intriguing transport phenomena
\cite{Park02,*Liang02,Gaudioso00,*Osorio10,*Riel06,*Choi06}.

Molecules can be considered as very small quantum dots.  An important aspect
that distinguishes them from semiconductor-based mesoscopic systems is the influence of the vibrational degrees of freedom. 
Due to the small size of molecules, the charging of the
molecular bridge is often accompanied by a significant change of the nuclear
geometry, which indicates strong coupling between electronic and vibrational
degrees of freedom.  It manifests itself 
in the current-voltage characteristics of molecular
junctions \cite{Secker11,Cuevas10}\cite{Smit02,*Natelson04,*Ballmann10} and may result
in current-induced heating of the molecular bridge
\cite{Huang06,*Natelson08,*Ioffe08} and large fluctuations \cite{Secker11}.
Conformational changes of the geometry of the conducting molecule are possible
mechanisms for switching behavior and negative differential resistance
\cite{Gaudioso00}.

The experimental findings have stimulated great interest in the basic
mechanisms of quantum transport at the molecular scale, in particular
effects due to electronic-vibrational coupling.  A variety of different
theoretical methods have been applied to study these phenomena, including scattering theory, nonequilibrium
Green's function approaches, and master equation methods (see, e.~g., \cite{Galperin07,Cuevas10} and references therein).  Although much
physical insight has been obtained by the application of these methods, all
these approaches involve significant approximations.

To elucidate the detailed mechanisms and address the full complexity of the
nonequilibrium transport problem, advanced numerical techniques that do not
involve intrinsic approximation are required. Powerful methods that have
been proposed in this context are the diagrammatic Monte Carlo simulation
(diagMC) approach \cite{Muehlbacher08,*Werner2009,*Werner2011}, multilayer
multiconfiguration time-dependent Hartree method in second quantization
representation (ML-MCTDH-SQR) \cite{Wang09,Wang11} as well as standard quantum Monte Carlo and iterative path integral summation scheme \cite{PhysRevB.72.041301,PhysRevB.85.121408}.

Here, we employ the first two methods to address two important and
largely unsolved questions: (i) How is the steady state in a molecular quantum
dot established starting from a specified preparation? (ii) Does the steady
state depend on the initial preparation, e.g.\ on the initial occupation of the
dot?  In particular (ii), which is closely related to the
phenomena of bistability and hysteresis, has been discussed controversially
based on approximate methods
\cite{my,Alexandrov2003,Galperin2005,Dzhioev11}. 
On the other hand, there are a number of works in which hysteretic behavior was observed experimentally \cite{li:645,*SMLL:SMLL200600101,*Liljeroth31082007}.
Our findings indicate that,
for certain parameter regimes, a remarkable memory effect in the nonequilibrium
dynamics exists, which manifests itself in different steady states for
different initial preparations. We discuss the relation of this finding to
earlier predictions of bistability for the model investigated in
\cite{my,Alexandrov2003,Galperin2005} as well as to stochastic switching
observed in nanoelectromechanical systems
\cite{PhysRevLett.92.248302,Koch2005,Mozyrsky2006,Pistolesi2008,Nocera2011}.

In order to study vibrationally coupled electron transport in a molecular
quantum dot we employ the resonant tunneling model given by the Hamiltonian
\cite{Glazman1988,Galperin07,Haertle08},
\begin{equation} \label{Hamiltonian}
H
=
H_D
+
H_{LR}
+
H_\text{ph}
+
H_T
+
H_I \,,
\end{equation}
which describes a single electronic level of energy $\epsilon_D$ (we shall
consider a spinless system and use units in which $\hbar=e=k_B=1$
throughout), $H_D = \epsilon_D d^\dag d$, coupled to two noninteracting
fermionic reservoirs, $H_{LR} = \sum_{\alpha, k} \epsilon_{\alpha k}
a^\dag_{\alpha k} a_{\alpha k}$, describing the left and right ($\alpha = L,R$)
metallic electrodes with energy dispersion $\epsilon_{\alpha k}$, and a bosonic
part, $H_\text{ph} = \sum_{\kappa} \omega_\kappa b^\dag_\kappa b_\kappa$, which
models the vibrational degrees of freedom of the molecule within the harmonic
approximation employing 
normal modes with frequencies
$\omega_\kappa$.  The fermionic environments serve as charge reservoirs,
inducing a nonequilibrium current by virtue of the coupling between the leads
and the molecular quantum dot
\begin{equation} 
H_T
=
\sum_{\alpha,k} \left(t_{\alpha k} a^\dag_{\alpha k} d + \text{h.c.}\right) \,,
\end{equation}
where $t_{\alpha k}$ denotes the dot-lead coupling strength between the $k$th
electronic level of lead $\alpha$ and the molecule. Finally,
electronic-vibrational coupling is described by the interaction part
\begin{equation} 
H_I
=
d^\dag d \sum_\kappa \lambda_\kappa \left(b_\kappa + b^\dag_\kappa\right)
+
d^\dag d \sum_\kappa \frac{\lambda^2_\kappa}{\omega_\kappa}
\end{equation}
with coupling constants $\lambda_\kappa$. The last term is a counter term that
corresponds to the polaron shift $\sum_\kappa \lambda^2_\kappa/\omega_\kappa$.
The molecule-lead and electron-vibrational couplings are characterized by the
hybridization $\Gamma_\alpha(\epsilon) = 2\pi\sum_k |t_{\alpha k}|^2
\delta(\epsilon - \epsilon_{\alpha k}) $ and spectral density $J(\omega) =
\pi\sum_\kappa \lambda^2_\kappa \delta(\omega - \omega_\kappa)$,
respectively. In our numerical studies we use different models for the
molecule-lead coupling; on the qualitative level the phenomena we observe are
universal and independent of them.

\begin{figure}
\begin{center}
   \includegraphics[width=8cm]{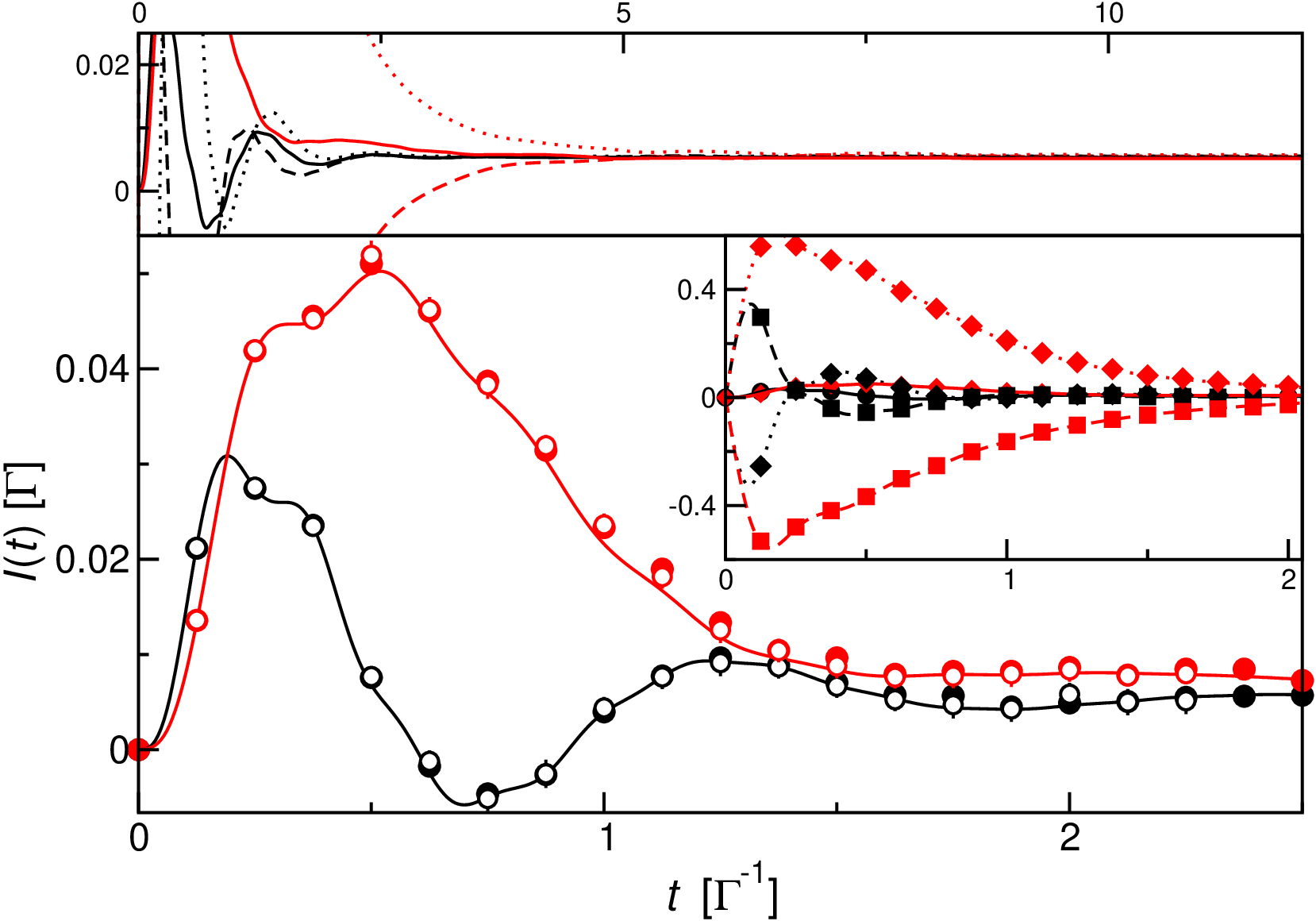}
   \caption{(color online) Comparison of ML-MCTDH-SQR (lines) and diagMC data (symbols) for an
initially empty (black) and occupied (red/grey) dot for $I(t)$ (circles/solid lines), $I_L(t)$ (squares/dashed lines), and $-I_R(t)$ (diamonds/dotted lines). 
The main graph shows the average current for ${\epsilon}_D = 4.7\Gamma$, 
$V = 1.25\Gamma$, $\Gamma_\alpha(\epsilon) = \Gamma \sqrt{1 - 
(\epsilon/\epsilon_c)^2}$ for $|\epsilon| \le \epsilon_c$ with bandwidth 
$2\epsilon_c = 50\Gamma$ (semielliptic band), and an Ohmic spectral density 
$J(\omega) = 2\pi\omega\exp(-\omega/\omega_c)$ with $\omega_c = 0.775\Gamma$
at zero temperature; filled and open symbols refer to $I(t)$ and $[I_L(t) - 
I_R(t)]/2$, respectively. The inset displays the different timescales on
which $I(t)$, $I_L(t)$, and $I_R(t)$ reach the stationary regime, while the 
upper panel demonstrates that all currents converge to the same stationary 
value.     \label{fig: comparison no hysteresis}}
\end{center}
\end{figure}

To characterize the nonequilibrium dynamics of the quantum dot, we consider the
dot population, $P(t) = \langle d^\dag(t)\, d(t) \rangle = {\rm tr} \lbrace
\rho_0 d^\dag(t)\, d(t)\rbrace$, and the electrical current through the dot,
$I(t) = (1/2) (d/dt) \langle N_L(t) - N_R(t)\rangle=[I_{L}(t)-I_R(t)]/2$.  Thereby, $N_\alpha =
\sum_k a_{\alpha k}^\dag a_{\alpha k}$ measures the number of particles in
contact $\alpha$, $I_{\alpha}(t)$ are the currents from the individual electrodes to the dot, and the initial preparation is given by the density operator
$\rho_0$ at time $t=0$, which describes a factorizing initial state, with the
dot being either empty or occupied. Initially 
the leads are in thermal equilibrium with energies shifted according to the 
chemical potentials $\mu_\alpha$; alternatively, one could keep each of the
leads filled up to different chemical potentials. In the wide 
band limit, the stationary currents from these two initial conditions have 
negligible numerical differences \cite{Schmidt2008}.
The bias $V = \mu_L - \mu_R$ ensures that for sufficiently long times a stationary state is
reached not only for the average current $I(t)$ , but also for the currents from the individual electrodes to the dot $I_{\alpha}(t)$, which are equal in the limit $t \to \infty$\footnote{We have also closely monitored the charge conservation, which was satisfied with high numerical precision in \emph{all} simulation runs.}. In practice $I(t)$ converges to the steady state value much faster than $I_\alpha(t)$ though, see Fig.~\ref{fig: comparison no hysteresis} and \cite{Muehlbacher08,Wang09}.

The reached stationary nonequilibrium
state is, if ergodicity holds, supposed to be unique and independent of the
initial preparation of the system. Therefore, even though different
preparations might yield different transient dynamics, in the long-time limit
one expects to find the transport properties independent of the evolution
history. In this Letter we investigate the validity of this conjecture.

We employ two different numerical approaches, the ML-MCTDH-SQR
method and the diagMC approach (see \cite{Muehlbacher08,Wang09} for details).
Briefly, the ML-MCTDH-SQR method is a variational basis-set approach to study
quantum dynamics for large systems containing identical particles.  Within this
approach the wave function is represented by a recursive, layered expansion in
Fock space employing the occupation number basis. Its time evolution is then
determined from the Dirac--Frenkel variational principle by dynamically
optimizing all the parameters.  On the other hand, the diagMC method relies on
an expansion of the time evolution in terms of the dot-lead coupling. After
integrating out all environmental degrees of freedom, one obtains an infinite
sum over Feynman diagrams which is then evaluated by a stochastic MC
scheme. Both approaches are \textit{numerically exact}, as the respective
errors can be made arbitrarily small: for diagMC, the statistical error due to
the MC sampling can be easily minimized by increasing the number of MC
measurements, while ML-MCTDH-SQR results can be systematically converged by
increasing the number of states as well as the basis set.
Both approaches yield consistent results for the transient dynamics as well as
for the stationary state. Fig.~\ref{fig: comparison no hysteresis} depicts
results of both approaches
 for the
time-dependent current for a model with a semielliptic molecule-lead coupling
and an Ohmic spectral density, showing excellent agreement.

While the results of Fig.~\ref{fig: comparison no hysteresis}, corresponding to
off-resonant transport, are an example for a stationary state that is
independent of the initial preparation, various claims have been made regarding the existence of bistability or hysteresis in the model under
investigation~\cite{my,Alexandrov2003,Galperin2005,DAmico08}. These claims,
however, received considerable criticism; the notion of bistable regimes in
this model is anything but generally accepted \cite{Mitra2004}. Usually, the
reasoning in favor of such effects assumes the existence of a parameter regime
(typically requiring strong electron-phonon coupling) within which the
effective potential of the phonon mode becomes multistable even in the
stationary regime. Accordingly, upon entering this regime from the smaller or
larger voltage domain, the system ends up in different potential minima,
resulting in different stationary states and corresponding different transport
properties for otherwise identical parameters \cite{my}.

These claims have been debated for almost a decade without reaching an accepted
conclusion. On the one side, a general proof of non-existence of bistability is
hard to give, while on the other side all previous works in favor of the
phenomenon rely on approximative methods (e.~g. the Born-Oppenheimer
approximation of \cite{my}) whose accuracy is difficult to judge. With our
numerically exact approaches at hand, however, we can give a more detailed
account of the underlying dynamics in the proposed regimes of bistability.
According to \cite{my} the two points in parameter space which minimize
the action \emph{and} result in two different values for the current correspond
to a nearly empty and a nearly fully occupied dot.  Thus, numerical simulations
with initial occupation numbers 0 and 1 should lead to the different dot
configurations required for bistability to occur. Therefore, in the following
we investigate the influence of the initial preparation on the stationary
state, which, in the absence of bistability, should not exist.
Since, as noted above, in 
the longtime limit all currents $I_\alpha(t)$, $I(t)$ converge to the same 
value, here we restrict ourselves to the stationary behavior of $I(t)$.

\begin{figure}
\begin{center}
   \includegraphics[width=8cm]{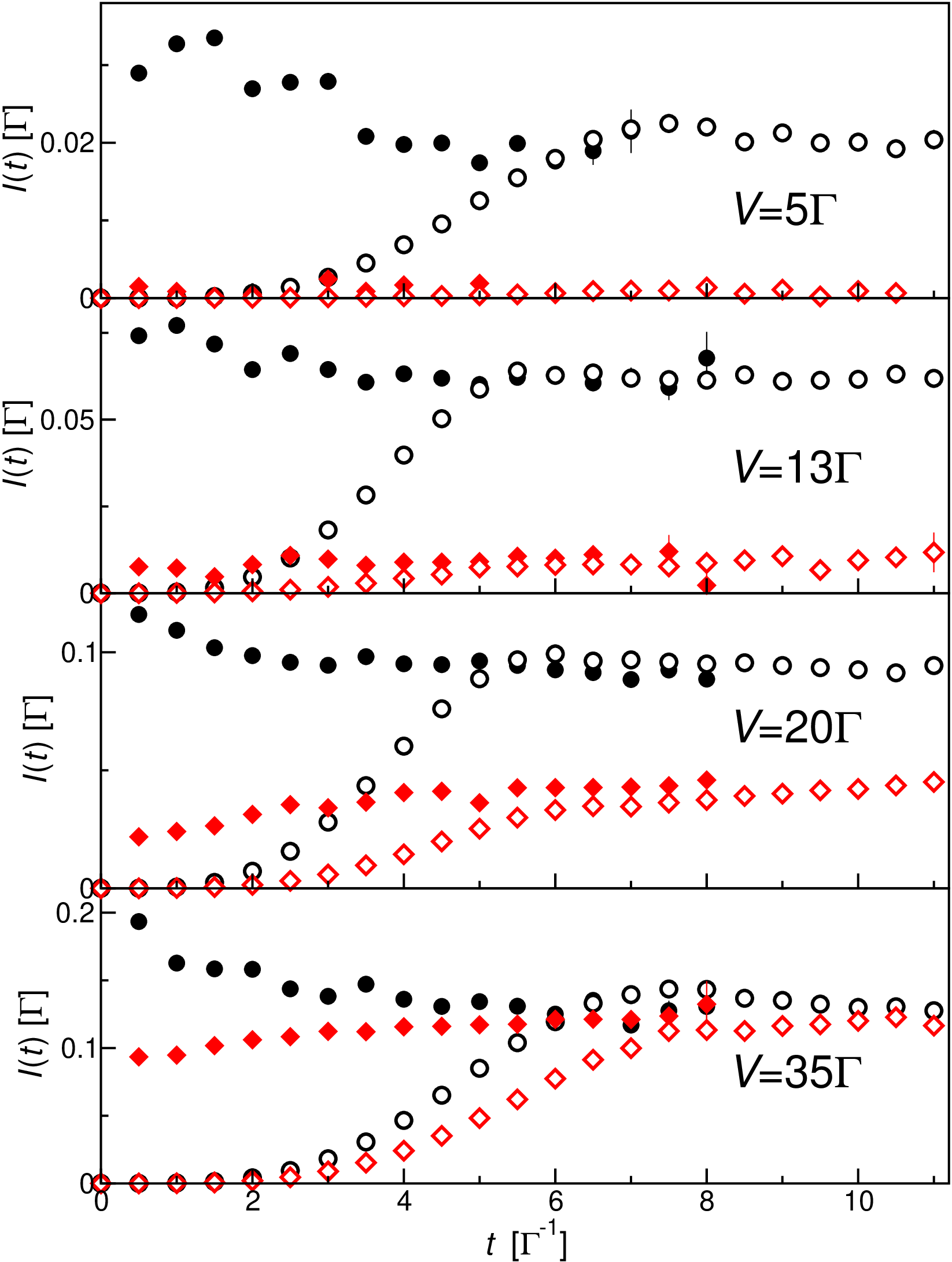}
   \caption{Current $I(t)$ for ${\epsilon}_D = -5\Gamma$,
     $\Gamma_\alpha(\epsilon) = \Gamma/2$ for $|\epsilon| \le \epsilon_c$ (flat band), and a single phonon mode
     with $\lambda_0 = 8\Gamma$ and $\omega_0 = 4\Gamma$, for an initially
     empty (black) and occupied (red) dot at zero temperature. Open symbols refer to diagMC data
     for a smooth switch-on of the tunneling coupling, $\Gamma(\epsilon, t) =
     \sin^2[\pi t/(2\tau_\text{sw})]\, \Gamma(\epsilon)$ for $t \le
     \tau_\text{sw}$ ($= 8\Gamma^{-1}$ for $V = 5\Gamma$ and $35\Gamma$, else $6\Gamma^{-1}$), while filled
     symbols refer to an instantaneous switch-on. The bandwidth is $2\epsilon_c
     = 41\Gamma$ for $V = 35\Gamma$, else $26\Gamma$.
     \label{fig: I bistable 1}}
\end{center}
\end{figure}

\begin{figure}
    \begin{center}
      \includegraphics[width=8cm]{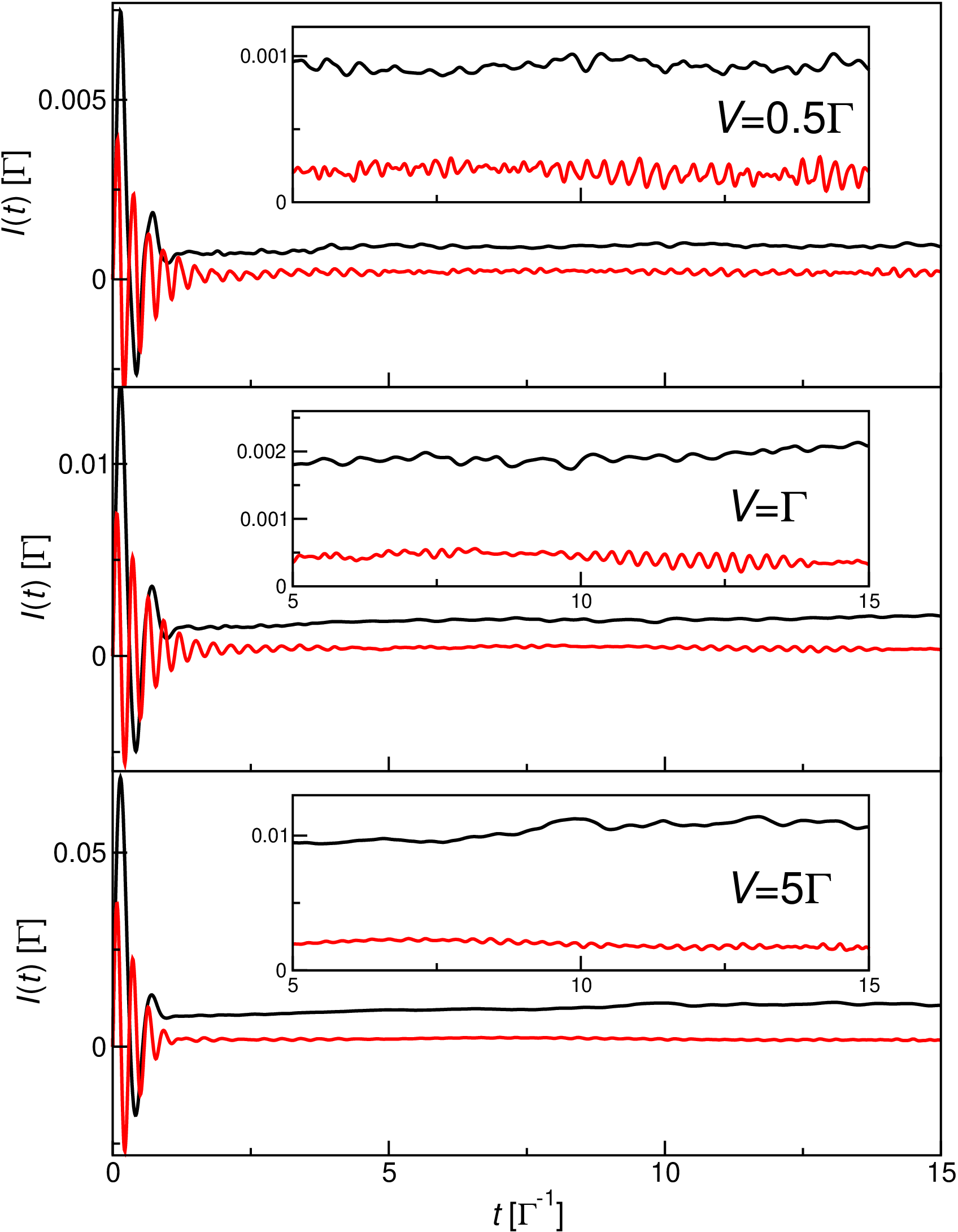}
      \caption{Similar to Fig.~\ref{fig: I bistable 1} but with ML-MCTDH-SQR
        result for the parameters: ${\epsilon}_D = -2.5\Gamma$, $\omega_0 =
        \Gamma/8$, $\lambda_0 = \Gamma$, and a bandwidth $2\epsilon_c =
        20\Gamma$. \emph{Insets:} Magnified versions of the same plots in the
        stationary regime.\label{fig: slow_mode} }
    \end{center}
\end{figure}

We consider a model with a single, strongly-coupled vibrational
mode, $J(\omega) = \pi \lambda_0^2 \delta(\omega - \omega_0)$.  Fig.~\ref{fig:
  I bistable 1} shows the time-dependent current for the two different initial
preparations (i.e.~empty and occupied dot) at different voltages for parameters
in the nonadiabatic regime, i.~e. $\omega_0 > \Gamma$.  Strikingly, the
currents converge towards stationary values on timescales roughly of the order
of $\Gamma^{-1}$ for almost all voltages.  This is in accordance with many
previous studies of the real-time dynamics, both for the
present~\cite{Riwar2009} as well as for the Anderson model~\cite{Schmidt2008}
and can be rationalized by noting that the strength of the tunneling coupling $\Gamma$ defines the typical timescale for charge dynamics.  However,
except for the high voltages the stationary values of the current are clearly
different, with the initially empty dot leading to a significantly larger
current than the initially occupied one. This intriguing finding contradicts
the notion of an unique stationary state and instead supports the idea of
bistability. A similar result is obtained in the adiabatic regime
of rather slow vibrational motion, i.e.\ $\omega_0 < \Gamma$, depicted in Fig.\ref{fig: slow_mode}.  This is the regime originally suggested for the
existence of bistability in the theoretical approaches \cite{my}. The numerical results in Figs.\ref{fig: I bistable 1},
\ref{fig: slow_mode} indicate the dependence of the long-time current on the
initial occupation of the dot. This finding is corroborated by the dynamics of
the population of the dot (data not shown). The population of the dot may
converge to its stationary value on a significantly longer timescale than the
current. However, for the examples considered above its stationary state
exhibits a dependence on the initial occupation even more pronounced than for
the current.

The comparison of the different panels in Fig.~\ref{fig: I bistable 1} shows
that the sensitivity of the dependence of the current on the initial
preparation is influenced by the applied bias voltage.  While for $V=35\Gamma$
both currents are approaching each other within a timescale $\tau_\text{st}$ of
the order of $\Gamma^{-1}$ they still seem to be transient for $V=20 \Gamma$
within the times accessible by our simulations. Here, it is not clear whether
both currents converge to different or the same stationary value. In the latter
case the corresponding stationary timescale $\tau_\text{st}$ would be much
larger than $\Gamma^{-1}$, offering an alternative explanation for the mismatch
in the current for not too large voltages in both, Figs.~\ref{fig: I bistable 1}
and \ref{fig: slow_mode}. Instead of identifying the plateau value as the
actual steady state, it could be interpreted as a transient current with a
timescale $\tau_\text{st}$ exceeding $\Gamma^{-1}$ by orders of magnitude.

The temperature $T$ of the leads has a similar influence on the effect: an
increase of $T$ destroys the bistability signatures just as an increase of $V$
does. On the other hand, it is noted that the difference in the long-time
behavior is independent of the details of the initial switch-on of the
tunneling coupling (cf.~Fig.~\ref{fig: I bistable 1}) as well as of the
specific form of the hybridization (i.e.~semielliptic vs.~flat band).

We would like to stress that the numerically exact, time-dependent methods
employed in our studies cannot directly address the stationary state, which is
formally obtained in the infinite time limit, but only map the dynamics over
long but finite timescales. It is thus not possible to unambiguously determine
whether the differences in the long-time transport behavior account for truly
distinct stationary states, or whether the corresponding transport properties
would still decay to the same unique stationary value, void of any imprints of
the preparation, on timescales significantly longer than those covered by our
calculations.  While in the latter case the concept of bistability does not
have to be introduced, an explanation for this surprisingly strong separation
of timescales is, to our knowledge, still lacking -- that is, the timescales
necessary to reach the true stationary state within and outside the
aforementioned parameter regime.

Since the pronounced transient dynamics always die out on the same timescale
(of the order of $\Gamma^{-1}$), we conclude that the effect is not
\textit{caused} but merely \textit{triggered} by the changes in the initial
preparations. Furthermore, the phenomenon seems to be completely unaffected by
details of the transient dynamics. For voltages at which we observe the
separation of timescales, the current originating from an instantaneous initial
switch-on of the tunneling coupling exhibits very rich transient dynamics,
suggesting a rather extensive exploration of the effective potential surfaces,
while for a continuous switch-on process, one finds quite a smooth convergence
process. Yet in both cases we observe the same separation of timescales.  Since
the system lacks any energy scale which can be set in relation to this novel
long timescale it should, if finite, be generated in a yet unknown way. 

We also would like to mention that the described phenomenon is not directly
related to the stochastic switching processes leading among other things to
random telegraph noise frequently discussed in the context of shuttling in
nanoelectromechanical devices
\cite{PhysRevLett.92.248302,Koch2005,Mozyrsky2006,Pistolesi2008,Nocera2011}.  While the
switching times would be of the order of $10^3-10^4$ in units of $\Gamma^{-1}$
for most of the plots shown above (we use the procedure from
\cite{Mozyrsky2006}) and thus apparently in accordance with the reported
data, we would like to point out that our simulations would then describe the
very \emph{onset} of the stochastic switching process, which is not covered by
any of the existing studies.

We finally comment on the applicability of our findings, obtained for the 
standard  model of vibrationally-coupled electron transport, to realistic
molecular junctions. 
Typical values of the coupling $\Gamma$ 
vary between meV and a few eV, depending on the specifics of the binding
group and the geometry. Assuming an average value of $\Gamma = 0.1$ eV (which is
realized, e.~g. in benzenealkenethiol-gold junctions \cite{doi:10.1021/jp711940n}), 
our results  predict the
phenomenon of bistability to occur both in the nonadiabatic regime (e.~g. for a
vibrational mode with a higher frequency  of $\omega_0 = 0.4$ eV as in Fig.~\ref{fig:  I bistable 1}) 
as well as in the adiabatic regime (e.~g.\ $\omega_0 = 0.0125$ eV as in Fig.~\ref{fig:
  slow_mode}) for sufficiently large electronic vibrational coupling
($\lambda/\omega_0 \gtrsim 1 $) and not too large voltages ($V < 2 $ V). These
parameters are expected to be of relevance for many molecular junctions,
e.g. for some of the molecular junctions, where hysteretic behavior was
observed experimentally \cite{li:645,*SMLL:SMLL200600101,*Liljeroth31082007}. According to our results, the phenomenon of
bistability persists on a timescale of at least a picosecond, which is longer than
typical vibrational periods. It should
be observable with methods of femtosecond spectroscopy. In view of
the current experimental progress \cite{Huang06,*Natelson08,*Ioffe08}, 
spectrosopic techniques that allow to directly monitor ultrafast processes in
molecular junctions are expected to become available in the near future.

To summarize, we have presented numerically exact results for the
nonequilibrium dynamics of a molecular quantum dot. In a wide parameter range we find a distinct dependence of the
steady state current on the initial preparation of the molecular junction. Our
analysis shows that this phenomenon is related to the previously predicted
bistability of the model. In contrast to the previous approximate approaches, the present
numerically exact results unambiguously prove the existence of bistability
signatures in the dynamics over a time scale significantly longer than the
tunneling time.

KFA acknowledges financial support from the European Research Area (ERA)
NanoScience (Project CHENAMON) and computational resources from the bwGRiD
project. HW acknowledges the support from the National Science Foundation
(CHE-1012479). MT acknowledges support by the DFG and thanks R.\ H\"artle for
insightful discussions.  LM acknowledges computational resources from the Black
Forest Grid Initiative. AK is supported by the DFG Grant No. KO-2235/3-1 and
CQD of the University of Heidelberg.

\bibliographystyle{apsrevM}
\bibliography{paper,thoss}

\end{document}